\documentclass[10pt, conference]{IEEEtran}
\IEEEoverridecommandlockouts
\usepackage{cite}
\usepackage{amsmath,amssymb,amsfonts}
\usepackage{graphicx}
\usepackage{textcomp}
\usepackage{xcolor}
\def\BibTeX{{\rm B\kern-.05em{\sc i\kern-.025em b}\kern-.08em
    T\kern-.1667em\lower.7ex\hbox{E}\kern-.125emX}}
\usepackage[utf8]{inputenc} 
\usepackage[T1]{fontenc}    
\usepackage{hyperref}       
\usepackage{url}            
\usepackage{booktabs}       
\usepackage{amsfonts}       
\usepackage{nicefrac}       
\usepackage{microtype}      
\usepackage{xcolor}         
\usepackage{amsmath}
\usepackage[capitalize]{cleveref}
\usepackage{algorithm,algpseudocode}
\usepackage{graphicx}
\usepackage{subcaption}
\usepackage{listings}
\usepackage{xcolor}
\usepackage{xspace}
\usepackage{multirow}
\usepackage{pgfplots}
\usepackage{pifont}
\usepackage{makecell}
\usepackage{colortbl}

\usepackage{authblk}

\pgfplotsset{compat=1.16}
\usepgfplotslibrary{groupplots}
\usetikzlibrary{positioning, arrows.meta, fit, backgrounds, calc, shapes.geometric, shapes.misc, decorations.pathreplacing}

\makeatletter
\@ifundefined{AddToHook}{}{%
  \AddToHook{cmd/appendix/before}{\def\cref@section@alias{appendix}}%
}
\makeatother

\definecolor{codegreen}{rgb}{0.0, 0.5, 0.0}
\definecolor{codegray}{rgb}{0.5, 0.5, 0.5}
\definecolor{codepurple}{rgb}{0.58, 0.0, 0.83}
\definecolor{codeblue}{rgb}{0.13, 0.29, 0.53}
\definecolor{backcolour}{rgb}{0.96, 0.96, 0.96}
\definecolor{commentgreen}{rgb}{0,0.5,0}
\definecolor{keywordblue}{rgb}{0,0,0.7}
\definecolor{stringred}{rgb}{0.6,0,0}
\definecolor{numberpurple}{rgb}{0.5,0,0.5}

\definecolor{blackborder}{RGB}{30,30,30}
\definecolor{pfprof}{RGB}{216,118,46}   \definecolor{pfprofbg}{RGB}{252,238,225}
\definecolor{pfagent}{RGB}{58,110,184}  \definecolor{pfagentbg}{RGB}{228,237,250}
\definecolor{pfvalid}{RGB}{63,163,77}   \definecolor{pfvalidbg}{RGB}{227,245,229}
\definecolor{pfsel}{RGB}{124,92,191}    \definecolor{pfselbg}{RGB}{238,232,250}
\definecolor{pfgray}{RGB}{130,130,130}  \definecolor{pfgraybg}{RGB}{238,238,238}

\lstdefinestyle{pythonstyle}{
    backgroundcolor=\color{backcolour},
    commentstyle=\color{codegreen}\itshape,
    keywordstyle=\color{codeblue}\bfseries,
    numberstyle=\tiny\color{codegray},
    stringstyle=\color{codepurple},
    basicstyle=\ttfamily\footnotesize,
    breakatwhitespace=false,
    breaklines=true,
    captionpos=b,
    keepspaces=true,
    numbers=left,
    numbersep=8pt,
    showspaces=false,
    showstringspaces=false,
    showtabs=false,
    tabsize=4,
    frame=single,
    rulecolor=\color{gray!50},
    language=Python,
    emph={np, ndarray},
    emphstyle=\color{teal},
    escapeinside={(*@}{@*)},
}

\newcommand{\sys}{PerfAgent\xspace}

\usepackage[T1]{fontenc}
\usepackage{mathpazo}
\usepackage{courier}
\usepackage{xcolor}
\usepackage[most]{tcolorbox}
\usepackage{enumitem}
\usepackage{fancyvrb}
\usepackage{inconsolata}

\setlist[enumerate]{noitemsep,topsep=0pt,partopsep=3pt,leftmargin=*}
\setlist[itemize]{noitemsep,topsep=0pt,partopsep=3pt,leftmargin=*}

\begin{document}

\title{\sys: Profiler-Guided Iterative Refinement for Repository-Level Code Optimization}


\renewcommand\Authfont{\bfseries}
\setlength{\affilsep}{0em}
\author[1]{%
    Ryan Deng
}
\author[2]{
    Yuanzhe Liu
}
\author[3]{
    Bastian Lipka
}
\author[2]{
    Yao Ma
}
\author[4]{
    Xuhao Chen
}
\author[1$\dagger$]{
    \\ Tim Kaler
}
\author[5$\dagger$]{
    Jatin Ganhotra
}

\affil[1]{Massachusetts Institute of Technology, Cambridge, Massachusetts, USA}
\affil[2]{Rensselaer Polytechnic Institute, Troy, New York, USA}
\affil[3]{IBM, Ehningen, Germany}
\affil[4]{Michigan State University, Lansing, Michigan, USA}
\affil[5]{IBM, Thomas J. Watson Research Center, Yorktown Heights, New York, USA}

\maketitle

\begingroup
\renewcommand\thefootnote{}\footnotetext{$^\dagger$Co-senior authors.}%
\addtocounter{footnote}{-1}\endgroup

\begin{abstract}

Large language model (LLM) agents now perform well on correctness-oriented repository-level tasks, including SWE-Bench issue resolution and feature implementation in real codebases. 
However, they still struggle with repository-level code optimization, which requires preserving behavior while improving runtime performance. Passing tests is not enough in this setting; a patch must preserve behavior, implement code optimization, and approach expert speedups. Current agents often miss bottlenecks hidden behind abstraction layers and native extensions, stop after shallow speedups, or insufficiently test the code patches that thus may silently break edge cases. We present PerfAgent, a profiler-guided, verifier-in-the-loop workflow that gives an off-the-shelf coding agent the feedback needed to find real hotspots, improve beyond the first passing patch, and use profiler evidence rather than timing alone to decide what to optimize next. On two challenging optimization benchmarks, GSO and SWE-fficiency-Lite, PerfAgent more than doubles the rate of expert-matching patches over OpenHands with GPT-5.1, improving from 19.6\% to 39.2\% on GSO and from 26\% to 74\% on SWE-fficiency-Lite. It also surpasses an oracle best-of-five baseline at substantially lower cost, showing that the gains come from better feedback rather than additional test-time sampling.
\end{abstract}


\section{Introduction}
LLM agents have shown remarkable capability in solving real-world software engineering tasks. Frontier LLMs have achieved strong results on benchmarks such as SWE-Bench, SWE-Bench Pro and Multi-SWE-Bench \cite{jimenez2024swebench, deng2025swebenchproaiagents, zan2025multiswebench}, which evaluate agents on their ability to solve issues and implement features on a wide range of complex real-world code repositories. However, these benchmarks only focus on correctness and ignore the performance impact of the changes produced by the agent. 

Recent benchmarks such as GSO\cite{gso}, SWE-fficiency\cite{swefficiency} and SWE-Perf \cite{sweperf} require agents to optimize code in real-world repositories. 
{Unfortunately, LLM agents struggle to produce changes that match the performance of those made by human experts~\cite{gso}.}
In general, optimizing code within a repository is more challenging than optimizing individual compute kernels, since changes required to get better performance are often complex and span multiple languages and abstraction boundaries in the repository.

Specifically, by diving into the benchmarks, we identify \textit{three failure modes} that prevent LLM agents from matching human experts on repository-level code optimization.
First, agents do not use profiling tools to their full effect, and thus often miss the real bottlenecks, especially those hidden across
abstraction boundaries or inside native extensions, in the repository.
Second, agents often prematurely terminate optimizing after shallow speedups, leaving more performance improvement opportunities on the table.
Third, agents tend to test the changes narrowly, leading to patches that are fast on target workloads but may silently break edge cases and downstream code paths elsewhere.

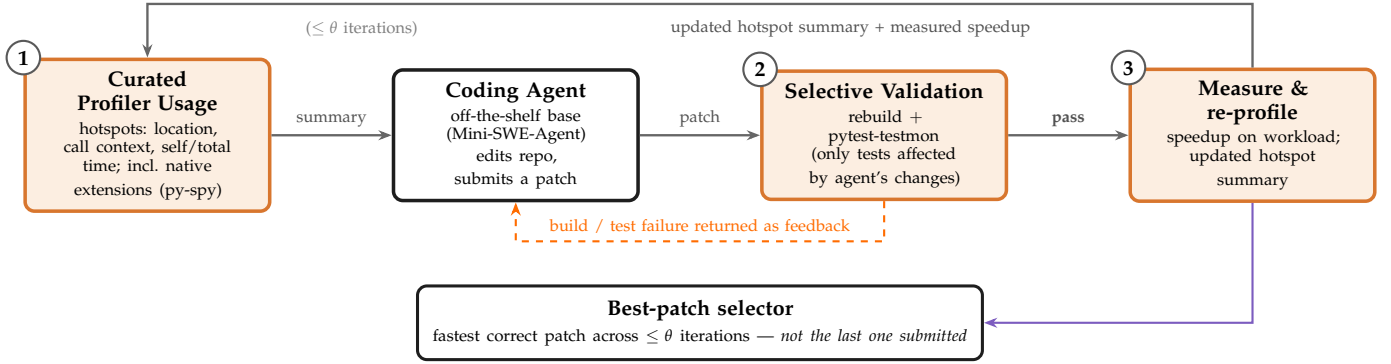
\begin{figure*}[t!]
\centering
\resizebox{\textwidth}{!}{%
\begin{tikzpicture}[
  font=\small,
  >={Stealth[length=2mm]},
  novbox/.style={rounded corners=4pt, draw, line width=1.5pt, align=center,
                 inner sep=5pt, minimum height=19mm, text width=33mm},
  basebox/.style={rounded corners=4pt, draw, line width=0.8pt, align=center,
                  inner sep=5pt, minimum height=19mm, text width=33mm},
  flow/.style={->, line width=1pt, draw=black!60},
  fb/.style={->, line width=0.9pt, draw=orange!85!red, dashed},
  lab/.style={font=\scriptsize, align=center, text=black!68},
  badge/.style={circle, draw=black!65, fill=white, line width=0.9pt,
                font=\bfseries\small, inner sep=1.5pt, minimum size=5mm},
]


\node[novbox, draw=pfprof, fill=pfprofbg] (prof) at (0,6)
  {\textbf{Curated Profiler Usage}\\[2pt]
   \scriptsize hotspots: location,\\
   \scriptsize call context, self/total\\
   \scriptsize time; incl.\ native\\
   \scriptsize extensions (py-spy)};

\node[novbox, draw=blackborder] (agent) at (5.5,6)
  {\textbf{Coding Agent}\\[2pt]
   \scriptsize off-the-shelf base\\
   \scriptsize (Mini-SWE-Agent)\\[1pt]
   \scriptsize edits repo,\\
   \scriptsize submits a patch};

\node[novbox, draw=pfprof, fill=pfprofbg] (valid) at (11,6)
  {\textbf{Selective Validation}\\[2pt]
   \scriptsize rebuild $+$\\
   \scriptsize pytest-testmon\\
   \scriptsize (only tests affected\\
   \scriptsize by agent's changes)};

\node[novbox, draw=pfprof, fill=pfprofbg] (meas) at (16.5,6)
  {\textbf{Measure \&}\\
   \textbf{re-profile}\\[2pt]
   \scriptsize speedup on workload;\\
   \scriptsize updated hotspot\\
   \scriptsize summary};

\draw[flow] (prof)  -- node[lab, above]{\scriptsize summary} (agent);
\draw[flow] (agent) -- node[lab, above]{patch} (valid);
\draw[flow] (valid) -- node[lab, above]{\textbf{pass}} (meas);

\draw[flow] (meas.north) -- ++(0,0.95) -| (prof.north);
\node[lab, anchor=south] at (10.5,7.35)
  {updated hotspot summary + measured speedup};
\node[lab, text=black!50, anchor=south] at (3.2,7.35)
  {($\le \theta$ iterations)};

\draw[fb] (valid.south) -- ++(0,-0.6) -| (agent.south);
\node[lab, text=orange!85!red, anchor=north] at (8.25,4.88)
  {build / test failure returned as feedback};

\node[
  rounded corners=4pt, draw=blackborder, line width=1.3pt,
  align=center, text width=82mm, minimum height=11mm,
] (sel) at (8.25,3.2) {
  \textbf{Best-patch selector}\\[1pt]
  \scriptsize fastest correct patch across $\le \theta$ iterations
  --- \emph{not the last one submitted}
};
\draw[->, line width=0.9pt, draw=pfsel] (meas.south) |- (sel.east);

\node[badge] at ($(prof.north west)+(-0.0,0.0)$)  {1};
\node[badge] at ($(valid.north west)+(-0.0,0.0)$)  {2};
\node[badge] at ($(meas.north west)+(-0.0,0.0)$)   {3};

\end{tikzpicture}%
}
\caption{Overview of the \sys workflow. Unlike a plain retry loop, \sys injects targeted feedback at three points in each iteration: \ding{172} a \textbf{\textit{curated profiler summary}} is fed to the agent before it begins, localizing hotspots across abstraction boundaries including native extensions; \ding{173} a \textbf{\textit{selective validation}} (pytest-testmon) stage that checks each submitted patch against only the tests affected by the agent's changes, catching correctness regressions cheaply without running the full suite; and after a passing patch, the \ding{174} \textbf{\textit{objective-driven controller}} re-profiles and returns the updated hotspot summary alongside the measured speedup, driving the next of up to $\theta$ iterations ($\theta$=5 in \cref{sec:eval}). Additionally, a \textit{best-patch selector} retains the fastest correct patch across all iterations rather than the last one submitted. \cref{fig:loop_controller_example} traces a concrete run of this loop on a GSO task.
}
\label{fig:overview}
\end{figure*}

We introduce \sys, a workflow for repository-level code optimization to specifically address the above challenges.
As shown in \cref{fig:overview}, \sys supplies the agent with targeted information at three stages of the agentic loop. 
First, it runs a \textit{sampling profiler} on the given workload and feeds the agent a curated summary of hotspots---their location, call context, and share of runtime---so the agent can localize where time is actually spent, including inside native extensions. 
Second, an \textit{objective-driven} loop controller refuses to accept the agent's first passing patch: after each submission it rebuilds, re-validates, and re-profiles the repository and asks the agent to keep optimizing, retaining the fastest correct patch across multiple attempts. 
Third, a \textit{selective test} stage runs the curated test cases affected by the agent's current changes, sufficiently catching any regressions with low cost and returning them as feedback. This guarantees that correctness is not forgotten during optimization. 

Together, these components turn a one-shot agent into a profiler-guided, verifier-in-the-loop optimizer. 
We implement \sys on top of Mini-SWE-Agent \footnote{\sys can be applied to any off-the-shelf coding agent such as SWE-Agent, Trae Agent or iSWE-Agent \cite{sweagent, trae_agent, iswe_agent}} as the base agent.
Experiments across both frontier and open-source models show that our workflow yields large and consistent gains over strong general-purpose SWE agents \cite{openhands, codex}. 
Specifically, with GPT-5.1, \sys raises the fraction of expert-matching patches by 2$\times$ on GSO (39.2\% vs.\ 19.6\%) and 2.8$\times$ on SWE-fficiency-Lite (74\% vs.\ 26\%). We also show that \sys can improve results for open-source models like Kimi-K2 by 1.5$\times$ on both GSO and SWE-fficiency-Lite.
Importantly, \sys is sample-efficient: it surpasses an \emph{oracle} best-of-five inference-scaling baseline at a fraction of the cost, evidence that the gains stem from better feedback, not merely more test-time compute. 

This paper makes the following contributions:

\begin{itemize}
    \item We characterize three failure modes that prevent general-purpose LLM agents from matching human experts on repository-level code optimization: (1) missing real bottlenecks, 
    (2) premature termination after the first passing patch, and (3) insufficient testing that leaves correctness regressions undetected.
    \item We present \sys, a profiler-guided, objective-driven agentic workflow for repository-level code optimization. {\sys features three dedicated mechanisms, i.e., curated profiler summary, selective test, and objective-driven controller, which specifically address our identified three failure modes.}
    \item We implement \sys and evaluation with various models shows that \sys more than doubles the rate of expert-matching patches over strong baselines ($2\times$ on GSO, $2.8\times$ on SWE-fficiency-Lite).
    \sys also beats an \emph{oracle} best-of-five baseline while spending far less, isolating profiler-guided feedback---rather than additional compute---as the source of the improvement.
\end{itemize} 

Note that we view performance engineering as complementary to existing agentic workflows, but it imposes a stricter requirement: whereas issue-resolution benchmarks ask only that a patch be correct, repository-level optimization demands a patch that both be correct \emph{and} can improve performance, judged against the optimization a human expert produced.

\section{Benchmarks Overview}
\label{sec:benchmarks_overview}
We evaluate \sys on GSO and SWE-fficiency \cite{gso, swefficiency}, two repository-level code optimization benchmarks.
{\cref{tab:benchmarks_overview} provides an overview of the benchmarks. }
Note that in this paper, we evaluate on SWE-fficiency-Lite, a randomly sampled subset of $100$ tasks from SWE-fficiency, which is used by \cite{swefficiency} when presenting the benchmark.

\begin{table*}
  \centering
  \caption{\small GSO and SWE-fficiency-Lite benchmark overview. It shows the median speedup achieved by the human expert baseline over the base repository, the median patch size (lines of code modified) of the human expert baseline, and the percentage of tasks in which the human expert baseline modified non-Python code such as C/C++ or Rust.
  *Python indicates Python-only changes in the human expert baseline patch; the remaining languages may be used in conjunction with Python.
  }
  \label{tab:benchmarks_overview}
  \small
  \setlength{\tabcolsep}{6pt}
  \begin{tabular}{@{}lrrcrrc@{}}
  \toprule
  Benchmark & \# Repos & \# Tasks & Speedup & Patch Size & \% Non-Py & \makecell{Language Breakdown \\{\small (Python*, C/C++, Cython, Rust)}}\\
  \midrule
  GSO & 10 & 102 & 2.43$\times$ & 140 & 59 & (41\%, 45\%, 10\%, 4\%) \\
  SWE-fficiency-Lite & 9 & 100 & 3.57$\times$ & 20 & 12 & (88\%, \ 1\%, 11\%, 0\%) \\
  \bottomrule
  \end{tabular}
\end{table*}

In code optimization benchmarks, the agent is given a code repository at a particular commit along with a script that shows an example workload involving the use of an API implemented in the repository. The agent is tasked with making changes to the repository to improve the performance of the script, and its changes are evaluated both in terms of correctness and performance. Correctness is validated on hidden tests that ensure the existing behavior of the API is preserved. Performance is measured against a human expert baseline by computing a speedup ratio $SR = \frac{\text{Speedup}_{\text{Agent}}}{\text{Speedup}_{\text{Human}}}$, where $\text{Speedup}_{\text{Agent}}$ and $\text{Speedup}_{\text{Human}}$ are the speedups achieved over the base repository by the agent's and the human expert's changes, respectively. The agent's final score on that task is a function of this ratio.

\begin{itemize}
    \item \textit{\textbf{GSO}} is composed of $102$ tasks spanned across $10$ popular python repositories such as numpy, pandas, and pydantic, and involves languages such as Python, Cython, C, C++ and Rust. In addition to the test script provided to the LLM agent, GSO also has other hidden tests that are used for evaluation. These hidden tests exercise the target API in diverse ways, and the benchmark uses them to assess both the correctness and performance of the agent's changes. The speedup ratio computed is averaged over all hidden tests.
    \item \textit{\textbf{SWE-fficiency-Lite}} is composed of $100$ tasks that span $9$ popular python repositories such as pandas, scikit-learn and sympy and involves languages such as Python, Cython, C and C++. Similar to GSO, SWE-fficiency provides a script to the LLM agent to optimize. Unlike GSO, however, it has no hidden tests. The provided test script alone is used for performance evaluation, while correctness is validated against a curated subset of the repository's existing test suite. 
\end{itemize}

\begin{figure*}[t!]
\centering
\includegraphics[width=\textwidth]{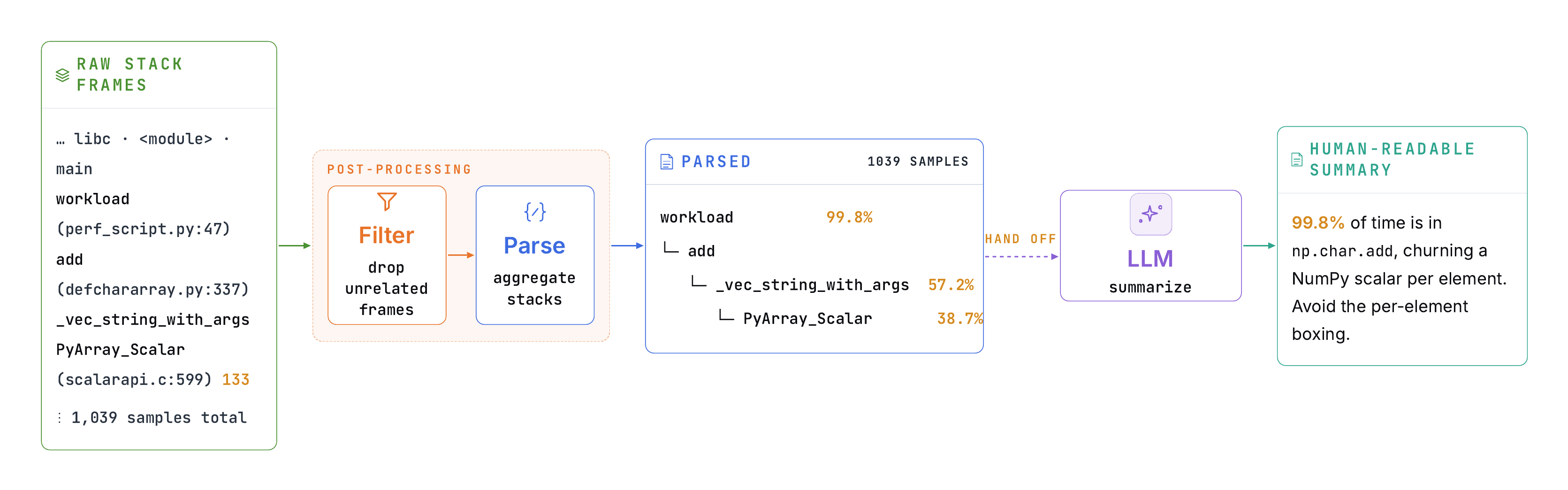}
\caption{Overview of \sys's profiler pipeline. Raw stack frames are collected by running py-spy on the workload, setup-related frames are filtered out, and the remaining samples are parsed and aggregated into hotspots with self-time, total-time, call context, and a native-vs-library flag. A separate LLM call then produces a concise natural-language summary that is passed to the agent.  In contrast, naive profiler usage --- running \texttt{cProfile} and forwarding its raw output --- yields noisy per-call counters covering only Python-layer frames, misses hotspots inside native extensions, and floods the agent's context window with irrelevant detail.
}

\label{fig:profiler}
\end{figure*}

    

\section{\sys's Workflow for Code Optimization}
\label{sec:workflow}

General-purpose LLM agents fail on repository-level code optimization in three characteristic ways:
\begin{itemize}
\item \textbf{Missing the real bottleneck.} Agents struggle to localize where time is actually spent: they often either skip profiling entirely or reach for Python's built-in \texttt{cProfile}, which carries high overhead and sees only Python-layer frames. As a result they fall back on surface-level Python edits and overlook hotspots hidden inside native extensions or beneath several layers of abstraction, yielding only modest speedups relative to a human expert's patch. On the NumPy string kernel of \cref{fig:loop_controller_example}, for example, \texttt{cProfile} cannot observe that 80\% of runtime is consumed by per-element C scalar allocation, so an unguided agent stops at the first Python-level fast path it finds at the original abstraction layer, whereas profiler feedback drives \sys to a purpose-built native C kernel ($5.87\times$); we trace this case end to end in \cref{sec:impact_control_loop}.
\item \textbf{Premature termination.} Once a patch passes the relevant tests and shows a noticeable speedup, the agent emits a stop signal, even when substantial headroom remains. Human experts, by contrast, have deep insight into how much improvement is still possible and when to stop, so an agent that settles for the first modest speedup performs far worse than an expert in practice.
\item \textbf{Insufficient testing of complex changes.} The changes that yield the largest speedups are often complex and reach into core library code. Yet coding agents only test the changes narrowly by running a limited range of tests. This results in a patch that is fast on the target workload but can silently break edge cases and downstream code paths elsewhere.
\end{itemize}
\sys introduces one targeted component to address each failure mode (\cref{sec:profiler,sec:controller,sec:testmon}).
\subsection{\sys's Workflow Overview}
\label{subsec:overview}

\cref{fig:overview} shows \sys's overall workflow. 
Each iteration of the loop consists of two phases: an inner agent phase and an outer evaluation phase.
In the inner phase, the agent is given the most recent profiler summary (see \cref{sec:profiler}) and runs inside our harness, until it submits a patch. The first iteration uses a profiler summary computed on the unmodified base repository and subsequent iterations use a summary computed on the repository with the agent's submitted patch from the previous iteration.

In the outer phase, the controller (detailed in \cref{sec:controller}) takes the submitted patch and applies it to the repository, rebuilds, and runs the curated test suite described in \cref{sec:testmon}. If the build fails or any test fails, we exit the outer phase early and return the error output to the agent as feedback for the next iteration. Otherwise, we run the profiler on the resulting repository with the agent's changes to produce a new summary that highlights the new hotspots, and pass it to the agent for its next attempt.

The controller allows the LLM up to $\theta$ submissions before it terminates. Across all iterations, we record the speedup of each submitted patch on the provided workload, and report the patch with the highest speedup as the final output. \sys is built on Mini-SWE-Agent \cite{sweagent} but is harness-agnostic: the same workflow can be layered onto any coding agent that emits a stop signal, such as OpenHands, Trae Agent, or iSWE-Agent \cite{openhands, trae_agent, iswe_agent}. 

\subsection{Hitting the Bottlenecks via Curated Profiler Usage}
\label{sec:profiler}

To address the first failure mode, \textit{missing the real bottleneck}, \sys runs a profiler on the workload to localize the code responsible for the runtime.
When general-purpose agents profile at all, they reach for Python's built-in cProfile, which as noted above, carries high overhead and sees only Python-layer frames. We instead use the py-spy sampling profiler, which has much lower overhead and profiles both Python code and native extensions \cite{pyspy}, removing both limitations; it interrupts execution at a fixed interval ($100$ times a second) and snapshots the call stack. 

However, a low-overhead, native-aware profiler is necessary but not sufficient: its raw output is voluminous, and forwarded as-is it would flood the agent's context with detail that buries the real bottleneck.
\sys therefore turns py-spy's raw samples into a compact, actionable summary (\cref{fig:profiler}) rather than forwarding them directly. We augment the provided workload script 
to run repeatedly for a fixed $10$-second profiling window (the while loop at line $15$) so that py-spy gathers enough samples, then discard samples related to setup such as reading or initializing data. We aggregate the rest into hotspots, each annotated with its location (filename, function name, line number), full call stack, share of samples, and an external-library flag that keeps the agent from editing code outside the repository. We report both self-time (excluding calls to other functions) and total-time (including them) so the agent can choose between speeding up a function and calling it less often. Finally, instead of dumping the full output into the prompt, a separate LLM call condenses it into a natural-language summary, bounding what the agent must read. We quantify the payoff of this curated usage over naive profiling in \cref{tab:ablation:profiler}.

Overall, we find that detailed profiling output helps agents pinpoint the parts of a repository responsible for performance overhead. This is especially valuable in large modern codebases where the many layers of abstraction can make it difficult for the agent to identify these hotspots by simply navigating the repository. We quantify how often this matters in \cref{sec:impact_control_loop}: \sys reaches low-level code (C, C++, Cython, or Rust) on $48\%$ of GSO instances versus $31\%$ for the OpenHands baseline, including $21$ instances where it goes low-level and OpenHands modifies only Python.

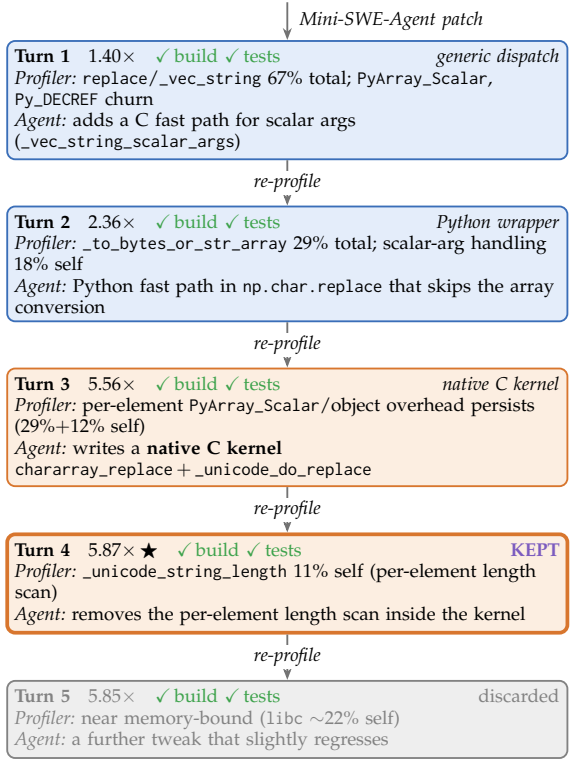
\begin{figure}[!b]
    \centering
    \begin{tikzpicture}[
      font=\scriptsize,
      >={Stealth[length=1.6mm]},
      row/.style={rounded corners=3pt, draw, line width=0.7pt, align=left, inner sep=3pt, text width=72mm},
      rp/.style={->, line width=0.7pt, draw=black!55},
    ]
    \node[row, draw=pfagent, fill=pfagentbg] (r1)
      {\textbf{Turn 1}\quad$1.40\times$\quad\textcolor{pfvalid}{\checkmark\,build \checkmark\,tests}\hfill\textit{generic dispatch}\\
       \emph{Profiler:} \texttt{replace}/\texttt{\_vec\_string} 67\% total; \texttt{PyArray\_Scalar}, \texttt{Py\_DECREF} churn\\
       \emph{Agent:} adds a C fast path for scalar args (\texttt{\_vec\_string\_scalar\_args})};
    \draw[rp] ([yshift=5mm]r1.north) -- node[midway,right=1pt,font=\scriptsize\itshape]{Mini-SWE-Agent patch} (r1.north);
    \node[row, draw=pfagent, fill=pfagentbg, below=6mm of r1] (r2)
      {\textbf{Turn 2}\quad$2.36\times$\quad\textcolor{pfvalid}{\checkmark\,build \checkmark\,tests}\hfill\textit{Python wrapper}\\
       \emph{Profiler:} \texttt{\_to\_bytes\_or\_str\_array} 29\% total; scalar-arg handling 18\% self\\
       \emph{Agent:} Python fast path in \texttt{np.char.replace} that skips the array conversion};
    \node[row, draw=pfprof, fill=pfprofbg, below=6mm of r2] (r3)
      {\textbf{Turn 3}\quad$5.56\times$\quad\textcolor{pfvalid}{\checkmark\,build \checkmark\,tests}\hfill\textit{native C kernel}\\
       \emph{Profiler:} per-element \texttt{PyArray\_Scalar}/object overhead persists (29\%$+$12\% self)\\
       \emph{Agent:} writes a \textbf{native C kernel} \texttt{chararray\_replace}\,$+$\,\texttt{\_unicode\_do\_replace}};
    \node[row, draw=pfprof, line width=1.4pt, fill=pfprofbg, below=6mm of r3] (r4)
      {\textbf{Turn 4}\quad$5.87\times$\,$\bigstar$\quad\textcolor{pfvalid}{\checkmark\,build \checkmark\,tests}\hfill\textbf{\textcolor{pfsel}{KEPT}}\\
       \emph{Profiler:} \texttt{\_unicode\_string\_length} 11\% self (per-element length scan)\\
       \emph{Agent:} removes the per-element length scan inside the kernel};
    \node[row, draw=pfgray, fill=pfgraybg, text=pfgray, below=6mm of r4] (r5)
      {\textbf{Turn 5}\quad$5.85\times$\quad\textcolor{pfvalid}{\checkmark\,build \checkmark\,tests}\hfill discarded\\
       \emph{Profiler:} near memory-bound (\texttt{libc} $\sim$22\% self)\\
       \emph{Agent:} a further tweak that slightly regresses};
    \draw[rp] (r1) -- node[midway,fill=white,inner sep=1pt,font=\scriptsize\itshape]{re-profile} (r2);
    \draw[rp] (r2) -- node[midway,fill=white,inner sep=1pt,font=\scriptsize\itshape]{re-profile} (r3);
    \draw[rp] (r3) -- node[midway,fill=white,inner sep=1pt,font=\scriptsize\itshape]{re-profile} (r4);
    \draw[rp] (r4) -- node[midway,fill=white,inner sep=1pt,font=\scriptsize\itshape]{re-profile} (r5);
    \end{tikzpicture}
    \caption{A concrete run of the \sys loop (\cref{fig:overview}) on the GSO task numpy\_\_numpy-1b861a2 (GPT-5.1) which optimizes the \texttt{numpy.char.replace} API.}
    \label{fig:loop_controller_example}
\end{figure}

\subsection{Objective-Driven Loop Controller}
\label{sec:controller}

To address the second failure mode, \textit{premature termination}, we push the agent past its first passing patch with an \emph{objective-driven} loop controller. Simply prompting the agent to keep going is not enough: without a concrete target to optimize toward, it tends to stop again after a marginal change, and without re-checking each new attempt, a faster but broken patch is accepted as progress. Our controller instead makes measured speedup a first-class objective and re-validates every attempt, so the agent keeps improving without sacrificing correctness. \sys's loop controller is an instance of verifier-in-the-loop self-refinement, in the lineage of Reflexion and Self-Refine \cite{reflexion, self_refine}, but specialized for performance: the feedback injected between attempts is concrete profiler and test output, and the objective being refined is measured speedup rather than correctness alone.\footnote{Iterating an agent toward an objective has recently become common in practice, e.g., the ``Ralph'' loop and scheduled agent re-invocation. Our contribution is not the loop itself but the optimization-specific feedback and objective that drive it.}


In Mini-SWE-Agent, when the agent is finished with its task, it emits a $\texttt{STOP}$ command that signals the harness to terminate the agentic loop. \sys's workflow intercepts this message and obtains the agent's submitted patch, but does not terminate the agentic loop. Instead, it applies the patch to the repository, builds the repository, runs tests and profiles the code, and the resulting feedback is returned to the agent who is then asked to continue optimizing based on the updated feedback on its latest attempt. We run this loop for up to $\theta$ iterations per task and report the patch with the highest measured speedup on the provided workload, rather than the last one submitted. This design pushes the agent past its natural stopping point without compromising correctness as every iteration re-validates the patch against the test suite, so a fast-but-broken submission is caught and surfaced as feedback rather than accepted. In addition, the controller helps the LLM agent refine its implementation, or even explore new directions, as the profiler gives immediate feedback on how each change performs.

In \cref{fig:loop_controller_example}, we trace how \sys's loop controller affects an agent's changes over each turn. The agent's submitted patch enters at the top; after each submission the controller returns the measured speedup, the outcome of rebuilding and running the selectively chosen tests, and a fresh profiler summary, and the agent then optimizes further. Updated profiler feedback drives the agent from Python-level fast paths (Turns~1--2) to a purpose-built native C kernel (Turn~3) and its refinement (Turn~4)---an optimization at a new abstraction layer. The controller retains the best patch (Turn~4, $5.87\times$), not the last (Turn~5, $5.85\times$, which regresses).

\begin{figure}[t]
\centering
\begin{tikzpicture}
\begin{groupplot}[
    group style={group size=1 by 2, vertical sep=0.75cm, xticklabels at=edge bottom},
    width=\columnwidth, height=3.4cm,
    xmin=0.8, xmax=5.2, xtick={1,2,3,4,5},
    tick label style={font=\footnotesize},
    label style={font=\footnotesize},
    legend style={font=\footnotesize, draw=none, fill=none, at={(0.02,0.96)}, anchor=north west},
    grid=major, grid style={gray!18},
]
\nextgroupplot[ylabel={Speedup ($\times$)}, ymin=1.3, ymax=7.7]
\addplot[thick, mark=*, color=blue!65!black] coordinates {(1,2.508)(2,3.707)(3,5.015)(4,5.727)(5,6.358)};
\addplot[thick, dashed, mark=square*, mark options={solid}, color=orange!90!black] coordinates {(1,1.697)(2,2.486)(3,2.638)(4,2.667)(5,2.672)};
\legend{GSO, SWE-fficiency-Lite}
\node[font=\scriptsize, anchor=south, yshift=2pt] at (axis cs:5,2.272) {$n{=}5$};
\nextgroupplot[xlabel={Controller turn}, ylabel={\# failures}, ymin=0, ymax=57, ytick={0,20,40}]
\addplot[thick, mark=*, color=blue!65!black] coordinates {(1,53)(2,37)(3,25)(4,14)(5,5)};
\addplot[thick, dashed, mark=square*, mark options={solid}, color=orange!90!black] coordinates {(1,14)(2,19)(3,11)(4,10)(5,3)};
\end{groupplot}
\end{tikzpicture}
\caption{Per-turn behavior of the loop controller on GSO (solid) and SWE-fficiency-Lite (dashed). \textbf{Top:} average (harmonic-mean) speedup of the patch submitted at each turn. \textbf{Bottom:} number of instances that hit a validation failure (build error or failing test) before submitting at that turn. The SWE-fficiency-Lite turn-5 average is computed over only $n{=}5$ instances, as most exhaust the per-task budget within one or two turns.}
\label{fig:per_turn}
\end{figure}
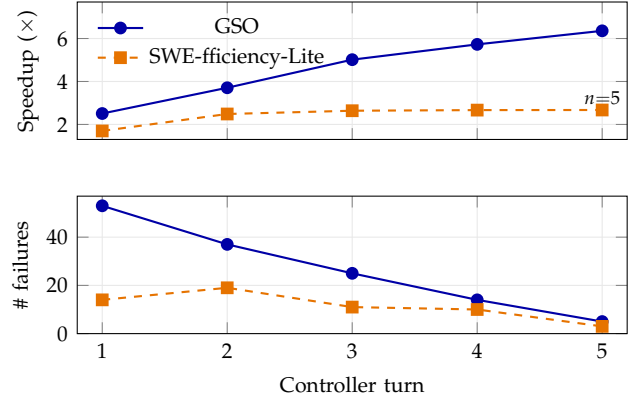

In addition, we see how the loop controller affects the speedup and correctness of an agent's changes over the course of $5$ turns in \cref{fig:per_turn}. We report the average (harmonic mean) speedup of the agent's submitted changes at each turn, along with the number of instances that encountered validation errors such as build errors and test failures before submission at that turn. Not all instances complete all $5$ turns, as some instances run out of budget before the full $5$ turns complete. Therefore, if an instance does not have a patch for a particular turn, we take the latest patch instead. 

On GSO, both quantities improve monotonically: the average speedup rises from $2.5\times$ at turn~1 to $6.4\times$ at turn~5, while the number of instances with validation failures falls from $53$ to $5$, showing that the controller extracts additional, correct speedup well past the agent's initial stopping point. On SWE-fficiency-Lite this trend is confounded by budget: most instances exhaust the per-task cost limit within one or two turns, so the speedup plateaus on later turns. In particular, \sys often spends many steps and tokens exploring the repository before writing any code. Therefore, with a $\$5$ budget, it is difficult to study the full impact of the control loop on an agent's changes in SWE-fficiency-Lite. We explore this in more detail in \cref{sec:eval:swefficiency_budget_per_task}. 

\subsection{Selective Validation against Correctness Regressions}
\label{sec:testmon}


{The third failure mode refers to the \textit{correctness regressions due to insufficient testing}. 
An alternative solution is to blindly include all the tests, but running the full test suite is prohibitively expensive on large real-world repositories.
To sufficiently catch any correctness regressions without causing significant overhead,
\sys's controller runs a curated test suite against the agent's patch after each iteration, and returns any failures as feedback. 
Specifically, we use pytest-testmon, a pytest plugin that uses code-coverage information to precisely select only the tests affected by the agent's changes, minimizing testing cost without missing any correctness regressions. }

Compared to running the repository's full test suite, our selective scheme reduces the number of tests run by $66\%$--$99\%$ on GSO and by $47\%$--$98\%$ on SWE-fficiency-Lite. {Details about the cost savings are described in Section B of the supplemental material. }
More importantly, we observe a clear increase in correctness of the final patch with our selective and curated test validation approach, in comparison to the ad-hoc testing done by the coding agent. We provide additional details in \cref{sec:eval}, with results in \cref{tab:main-results-gpt}.



\section{Evaluation}
\label{sec:eval}

Our evaluation is designed to test both the end-to-end effectiveness of \sys and the feedback mechanisms that explain its gains. 

We organize the main results around the following three questions. \textbf{RQ1}: does \sys improve repository-level optimization performance over strong agent baselines? \textbf{RQ2}: which parts of the workflow (profiler feedback, objective-driven iteration, and validation) account for the improvement? \textbf{RQ3}: how does iterative profiler feedback change the agent's optimization behavior? 

Further, we run analyses to rule out alternative explanations and characterize limitations: an oracle best@5 comparison tests whether the gains are merely due to more test-time sampling, open-source model results check whether the workflow transfers beyond GPT-5.1, reward-hacking analysis shows why optimization loops require robust measurement and hack-adjusted scoring, and a budget sensitivity analysis characterizes how the per-task cost cap constrains iteration on SWE-fficiency-Lite.

\subsection{Experimental Setup}

We evaluate \sys on GSO and SWE-fficiency-Lite using GPT 5.1, a frontier model, and Kimi-K2, an open-source model \cite{gpt, kimi}. For Kimi, on GSO we use Kimi-K2-0711 and on SWE-fficiency-Lite, we use Kimi-K2-0905, as those are the specific versions each benchmark evaluated on. Each benchmark provides patches produced by an agent implemented with OpenHands, which we evaluate to obtain baseline results \cite{openhands}. 


We implement \sys on top of Mini-SWE-Agent \cite{sweagent}, a lightweight single-agent harness with tool access. We use two variants depending on the model: for GPT-5.1 we use the default Mini-SWE-Agent with access to bash; for Kimi-K2 we augment it with a structured file editing tool from OpenHands \cite{openhands}, since open-source models struggle to view and edit files reliably through raw bash commands. When evaluating \sys, we run the controller loop for $\theta = 5$ iterations, set a maximum cost of $5$ dollars per task and a step limit of $200$ steps per task. The prompt used to evaluate \sys is in Section A of the supplemental material.

We run all agents in a separate AWS ec2 c6i.8xlarge instance which has $32$ CPUs and $64$GB of RAM. For evaluation, we use an AWS m8i.16xlarge which has $64$ CPUs and $256$GB of RAM, mirroring the evaluation setup described in GSO and SWE-fficiency \cite{gso, swefficiency}. When running GPT-5.1, we use high reasoning effort on all experiments, matching the experimental settings used in GSO and SWE-fficiency \cite{gso, swefficiency}. When running Kimi-K2, we use the default recommended settings \cite{kimi_k2, kimi_k2_2}. 

\subsubsection{Evaluation Metrics}
GSO and SWE-fficiency report different evaluation metrics based on the speedup ratios $SR = \frac{\text{Speedup}_{\text{Agent}}}{\text{Speedup}_{\text{Human}}}$ achieved on each task. GSO uses Opt@1: the percentage of tasks that are correct and have a speedup ratio of at least $0.95$. For SWE-fficiency, each task is assigned a score based on the speedup ratio. If the agent's changes do not pass correctness, then the score assigned is $\frac{1}{\text{Speedup}_{\text{Human}}}$. Otherwise, it is equal to the speedup ratio. The overall score for an LLM on the SWE-fficiency benchmark is the harmonic average of the scores for each task. While this metric rewards LLM agents for producing patches that surpass the performance of a human expert baseline, it can heavily skew the overall performance due to outliers in the benchmark. For example, producing an incorrect patch on a task in which the human expert baseline achieves a large speedup over the base repository is much more punishing compared to producing an incorrect patch on a task in which the human expert baseline achieves a small speedup. 

Therefore, for this paper, we report all benchmark results using GSO's Opt@1 score. In addition, we also report Correctness, the percentage of tasks that pass the benchmark's correctness tests, and Speedup@1 numbers (Sp@1), the percentage of tasks that achieve at least some speedup over the base repository. For Sp@1, we use a threshold of $1.2\times$ speedup, taken from GSO, as the threshold is large enough to remove patches that may achieve speedup solely due to timing variance \cite{gso}.

\subsubsection{Hack Detection}
\label{sec:eval:hack_detection}
GSO and SWE-fficiency both employ different complementary methods to prevent LLM agents from cheating on code-optimization benchmarks. GSO uses an LLM-as-a-judge to detect reward hacking, and SWE-fficiency identifies when LLM agents use stack-introspection to change behavior when it is being executed in a timing loop versus a correctness test.

We found that SWE-fficiency's detector, when used alone, missed common reward-hacking patterns. The most common hack we observed was caching results inside benchmark timing loops. SWE-fficiency's harness runs each evaluation in a separate Python process to prevent run-to-run caching of results, but within a single run timing is performed multiple times in a loop and is vulnerable to caching. Indeed, we observed many instances where agents exploited this structure to reduce reported benchmark timings. GSO's LLM-as-a-judge detector is better able to detect this style of reward hacking that is missed by SWE-fficiency's detector. However, SWE-fficiency's detector is still a useful addition to GSO's LLM judge because it is able to identify a unique form of reward hacking where an agent uses stack introspection to run different versions of a program during benchmarks and correctness tests.



We implemented a hack detector that combines the hack detection methods used by SWE-fficiency and GSO.
For both benchmarks, we report the results of running the combined hack detector as the hack-adjusted score.



\subsection{RQ1: End-to-End Optimization Performance}
\label{sec:eval:rq1}
We first address \textbf{RQ1}: does \sys improve repository-level optimization performance over strong agent baselines? We report end-to-end results against the OpenHands and Codex baselines (\cref{tab:main-results-gpt}), then break the comparison down task by task (\cref{tab:mcnemar}).

\begin{table}[t]
    \centering
    \footnotesize
    \setlength{\tabcolsep}{6pt}
    \caption{\small Results on the GSO and SWE-fficiency-Lite benchmarks for GPT-5.1. Correctness, Sp@1, Opt@1 score, and the hack-adjusted score are reported for each benchmark.}
    \label{tab:main-results-gpt}
    \begin{tabular}{lrrr @{\hskip\tabcolsep} r}
    \toprule
    Method & Correctness & Sp@1 & Opt@1 & Hack-Adj. \\
    \midrule
    \multicolumn{5}{l}{\textbf{GSO}} \\
    \addlinespace[1pt]
    OpenHands & 88.2 & 46.1 & 20.6 & 19.6 \\
    Codex & 89.2 & 48.0 & 18.6 & 17.7 \\
    $A_L$ (loop only) & 91.2 & 67.6 & 33.3 & 29.4 \\
    \quad + Tests & 96.1 & 47.1 & 24.5 & 20.6 \\
    \quad + Profiler & 95.1 & 70.6 & 36.3 & 34.4 \\
    \rowcolor{pfagentbg} \sys & 96.1 & \textbf{77.5} & \textbf{44.1} & \textbf{39.2} \\
    \midrule
    \multicolumn{5}{l}{\textbf{SWE-fficiency-Lite}} \\
    \addlinespace[1pt]
    OpenHands & 82 & 47 & 27 & 26 \\
    Codex & 80 & 59 & 39 & 39 \\
    $A_L$ (loop only) & 80 & 64 & 49 & 46 \\
    \quad + Tests & 93 & 64 & 49 & 46 \\
    \quad + Profiler & 83 & 73 & 59 & 57 \\
    \rowcolor{pfagentbg} \sys & 90 & \textbf{83} & \textbf{75} & \textbf{74} \\
    \bottomrule
    \end{tabular}
\end{table}

\textbf{GSO and SWE-fficiency-Lite Results.}
On GSO and SWE-fficiency-Lite, we compare \sys against the OpenHands baseline provided by each benchmark, as well as running on Codex, a frontier agentic harness developed by OpenAI \cite{codex}, and the results are shown in \cref{tab:main-results-gpt}. For each benchmark, we report the Opt@1 and hack-adjusted score. We see that on both benchmarks, \sys significantly outperforms both the OpenHands and Codex baselines. The Opt@1 scores on GSO are much lower compared to SWE-fficiency-Lite as GSO has hidden performance tests which evaluate the agent's changes on different workloads, while SWE-fficiency evaluates performance solely based on the provided script. Therefore, in GSO, the agent's changes may yield high speedups on the provided workload, but fail to generalize to unseen inputs. 

\textbf{Per-Task Win/Loss Analysis.}
Here we break our headline GPT-5.1 comparison down task by task, tabulating where \sys and the OpenHands baseline each match or exceed the human expert baseline on GSO and SWE-fficiency-Lite (\cref{tab:mcnemar}). Because both agents run on the same tasks, this is a \emph{paired} comparison: tasks where they agree---both optimize, or neither---say nothing about which agent is better, so significance comes entirely from the decisive tasks, where exactly one agent optimizes. We perform a more in-depth analysis of the instances in which the OpenHands baseline performs better in Section D of the supplemental material.

\begin{table}[t]
\centering
\footnotesize
\setlength{\tabcolsep}{8pt}
\caption{Per-task win/loss record of \sys against the OpenHands baseline. An agent \emph{wins} a task when it matches or exceeds the human expert baseline (and passes the hack detector) while the other does not; the remaining tasks are ties that both or neither agent optimizes.}
\label{tab:mcnemar}
\begin{tabular}{lrr}
\toprule
Per-task outcome & GSO & SWE-fficiency-Lite \\
\midrule
\rowcolor{pfagentbg} \sys wins & 27 & 52 \\
OpenHands wins & 7 & 4 \\
\addlinespace
Tie, both optimize & 13 & 22 \\
Tie, neither optimizes & 55 & 22 \\
\midrule
Total tasks & 102 & 100 \\
\bottomrule
\end{tabular}
\end{table}


\subsection{RQ2: Which Components Drive the Improvement}
\label{sec:eval:rq2}
\label{sec:ablation}
We now turn to \textbf{RQ2}: which parts of the workflow account for the improvement? We isolate the contribution of each component in \sys's loop controller by running several agents with \sys's controller for $5$ iterations. We evaluate on $3$ separate agents which we describe below.

\begin{itemize}
    \item $A_L$: An LLM agent with a stripped-down version of \sys's loop controller, where only timing feedback is given to the LLM at the end of each turn.
    \item \textit{$A_L$+Tests}: An LLM agent with \sys's loop controller that provides timing feedback and feedback from running the curated test suite.
    \item \textit{$A_L$+Profiler}: An LLM agent with \sys's loop controller that provides timing and profiler feedback. 
\end{itemize}

We compare these $3$ agents to \sys which provides a combination of timing feedback, profiler feedback, and feedback from running the curated test suite.

\cref{tab:main-results-gpt} provides the results of our comparison of these agents with the OpenHands baseline and \sys.  We find that $A_L$ outperforms both the OpenHands and Codex baselines in terms of Opt@1 and hack-adjusted scores, showing the value of test-time scaling with runtime feedback. The addition of correctness-test feedback in \textit{$A_L$+Tests} increases correctness on both benchmarks, but fails to improve Opt@1 or hack-adjusted scores on GSO and SWE-fficiency. In fact, on GSO the addition of correctness tests leads to markedly \emph{worse} performance on these metrics. The addition of profiling feedback in \textit{$A_L$+Profiler} improves Opt@1 and hack-adjusted scores on both benchmarks relative to $A_L$, but produces a correct patch less often than \textit{$A_L$+Tests}. The best Opt@1 and hack-adjusted scores are obtained by \sys which receives both profiler and correctness test feedback. The combination of both correctness test and profiler feedback in \sys achieves the best Opt@1 and hack-adjusted scores. \sys is able to pursue ambitious optimizations that are informed by profiling feedback, and the correctness test feedback helps it avoid correctness issues that arise from complex code changes.

\subsection{RQ3: Effect of Profiler Feedback}
\label{sec:eval:rq3}
We next study \textbf{RQ3}: how does iterative profiler feedback change the agent's optimization behavior? We first show in aggregate that profiler feedback pushes the agent across abstraction boundaries, then trace a single task through the loop as a case study.

\subsubsection{Optimization Across Abstraction Boundaries}
A key driver of these gains is that profiler feedback pushes the agent across abstraction boundaries and allows the agent to better understand the performance of native code. \sys modifies low-level code (C, C++, Cython, or Rust) on $48\%$ of instances, compared to $31\%$ for the OpenHands baseline, which leads to better results on these lower-level tasks. To further demonstrate this, we also report the results of GSO, broken down by Python-only tasks and the remaining non-Python-only tasks in \cref{tab:gso-lang-breakdown}. On GSO, $41.2\%$ of tasks in GSO involve a human expert baseline that only modifies Python code files, and on the remaining tasks, the human expert baseline additionally modifies low-level code (C, C++, Cython, Rust) files. Here, we see that agents perform worse on these low-level tasks compared to the Python-only tasks. However, on the agents that use a profiler, performance on the non-Python-only tasks significantly improves.\footnote{We do not perform the same breakdown on SWE-fficiency-Lite, because $89\%$ of its tasks involve Python-only changes.}


\begin{table}
    \centering
    \footnotesize
    \setlength{\tabcolsep}{6pt}
    \caption{GSO results broken down by language: 42 Python-only tasks vs. 60 non-Python-only tasks. Values reported are the percentage of tasks that match (or exceed) the human expert baseline and pass the hack detector.}
    \label{tab:gso-lang-breakdown}
    \begin{tabular}{lrr}
    \toprule
    Method & Python only & Non-Python-only \\
    \midrule
    OpenHands & 31.0 & 11.7 \\
    $A_L$ (loop only) & 45.2 & 18.3 \\
    \quad + Tests & 35.7 & 10.0 \\
    \quad + Profiler & 47.6 & 25.0 \\
    \rowcolor{pfagentbg} \sys & \textbf{50.0} & \textbf{31.7} \\
    \bottomrule
    \end{tabular}
\end{table}

\subsubsection{Case Study: Impact of \sys's Control Loop}
\label{sec:impact_control_loop}
We further expand on the example shown in \cref{fig:loop_controller_example} to demonstrate how \sys enables an agent to iteratively improve its implementation.

We look at task: \texttt{numpy\_\_numpy-1b861a2} from GSO which aims to optimize \texttt{numpy.char.replace} to demonstrate the impact of \sys's control loop on the agent's implementation. \cref{fig:loop_controller_example} traces how the agent's implementation evolves across the $5$ controller iterations. The initial profiler report attributes most of the runtime to the \texttt{replace} call site and the generic \texttt{\_vec\_string} dispatch path, and the agent responds with a C-level fast path for scalar arguments (Turn~1, $1.40\times$) and then a Python-level fast path that avoids an array-conversion step (Turn~2, $2.36\times$). After each attempt, the updated profiler feedback continues to report large overheads from per-element scalar creation (\texttt{PyArray\_Scalar}) and object reference counting that cannot be removed at the Python or generic-dispatch layer. This pushes the agent to implement a dedicated native kernel, \texttt{chararray\_replace}, in C (Turn~3), more than doubling the speedup to $5.56\times$. The feedback on the new kernel then localizes a per-element string-length scan (\texttt{\_unicode\_string\_length}), which the agent removes to reach its best patch (Turn~4, $5.87\times$). The final attempt explores a further change that slightly regresses (Turn~5, $5.85\times$), so the controller reports the Turn-4 patch rather than the last submission, illustrating why we retain the best measured patch instead of the final one. As in the aggregate trend (\cref{fig:per_turn}), the updated profiler feedback lets the agent discover an optimization at an entirely new abstraction layer, matching the approach taken by the human expert baseline. By contrast, the OpenHands baseline stops at the first fast path it finds at the original abstraction layer, as that already achieves speedup over the base repository. 


\begin{table}[t]
    \centering
    \footnotesize
    \setlength{\tabcolsep}{6pt}
    \caption{Results on the GSO and SWE-fficiency-Lite benchmarks for Kimi-K2. Correctness, Sp@1, Opt@1, and hack-adjusted scores are reported.}
    \label{tab:main-results-kimi}
    \begin{tabular}{lrrr @{\hskip\tabcolsep} r}
    \toprule
    Method & Correctness & Sp@1 & Opt@1 & Hack-Adj. \\
    \midrule
    \multicolumn{5}{l}{\textbf{GSO}} \\
    \addlinespace[1pt]
    OpenHands & 64.7 & 23.5 & 9.8 & 3.9 \\
    \rowcolor{pfagentbg} \sys & 89.2 & 31.4 & 10.8 & \textbf{5.9} \\
    \midrule
    \multicolumn{5}{l}{\textbf{SWE-fficiency-Lite}} \\
    \addlinespace[1pt]
    OpenHands & 74 & 39 & 26 & 22 \\
    \rowcolor{pfagentbg} \sys & 87 & 67 & 46 & \textbf{32} \\
    \bottomrule
    \end{tabular}
\end{table}

\subsection{Additional Analyses}
\label{sec:eval:additional}
Finally, we run additional analyses to rule out alternative explanations for \sys's gains and to characterize its limitations.

\subsubsection{Test-Time Scaling}
The improved performance of \sys is not merely due to its use of additional test-time compute. To demonstrate this, we evaluated \sys against a best-case test-time scaling approach that runs the OpenHands workflow 5 times and then uses an oracle to pick the best result. The use of an oracle to pick the best patch, in this setting, is the ``best-case'' for test-time scaling because it uses the results of hidden performance and correctness tests from the evaluation harness that are not made available to \sys or other baselines during inference. The results of this experiment are provided in \cref{tab:test-time-scaling}. On GSO, \sys achieves a better Opt@1 score than OpenHands best@5 (44.1\% vs 26.5\%) and a better hack-adjusted score (39.2\% vs 26.5\%) while costing over $3\times$ less. On SWE-fficiency-Lite, the results are similar with \sys achieving a better Opt@1 score (75\% vs 71\%) and hack-adjusted score (74\% vs 68\%) than OpenHands best@5 while costing $2\times$ less.



\begin{table}[t]
    \centering
    \footnotesize
    \setlength{\tabcolsep}{2.5pt}
    \caption{Comparison with test-time scaling running on OpenHands best@5 with an oracle judge. Correctness, Sp@1, Opt@1, Hack-adjusted score and cost in terms of average \$ per task are reported for each benchmark.}
    \label{tab:test-time-scaling}
    \begin{tabular}{lrrrrr}
    \toprule
    Method & Correctness & Sp@1 & Opt@1 & Hack-adj. & Cost (\$) \\
    \midrule
    \multicolumn{6}{l}{\textbf{GSO}} \\
    \addlinespace[1pt]
    OpenHands best@5 & 99.0 & 67.6 & 26.5 & 26.5 & 11.01 \\
    \rowcolor{pfagentbg} \sys & 96.1 & 77.5 & 44.1 & \textbf{39.2} & \textbf{2.88} \\
    \midrule
    \multicolumn{6}{l}{\textbf{SWE-fficiency-Lite}} \\
    \addlinespace[1pt]
    OpenHands best@5 & 99 & \textbf{93} & 71 & 68 & 9.91 \\
    \rowcolor{pfagentbg} \sys & 90 & 83 & 75 & \textbf{74} & \textbf{4.25} \\
    \bottomrule
    \end{tabular}
\end{table}

\subsubsection{Open-Source Models}
We also evaluate two open-source models: Kimi-K2-0711 on GSO, and Kimi-K2-0905 on SWE-fficiency-Lite, as those are the model versions used in the OpenHands baselines.

The results are shown in \cref{tab:main-results-kimi} and show that \sys outperforms the OpenHands baseline on both benchmarks. Overall, we find that open-source models generally struggle to identify the relevant information from the profiling output and implement the optimizations suggested from the profiler feedback. However, despite this, \sys still outperforms the OpenHands baselines on both benchmarks.

\subsubsection{Profiler Instruction in Prompt}
\label{tab:ablation:profiler}
An alternative to \sys's workflow that runs the profiler and provides a curated summary to the LLM is to give the agent instructions in the prompt on how to use the profiler, and have the agent decide when to use it. We separately evaluate an agent $A_P$ that is given instructions in the prompt detailing how \sys uses a profiler, and the results are shown in \cref{tab:use-of-a-profiler}.

\begin{table}
    \centering
    \caption{Results comparing \sys to an agent given instructions in the prompt on how to use a profiler.}
    \label{tab:use-of-a-profiler}
    \footnotesize
    \setlength{\tabcolsep}{4pt}
    \begin{tabular}{@{}lcccc@{}}
    \toprule
    Method & Correctness & Sp@1 & Opt@1 & Hack-Adj. \\
    \midrule
    \multicolumn{5}{l}{\textbf{GSO}} \\
    \midrule
    OpenHands & 88.2 & 46.1 & 20.6 & 19.6 \\
    $A_P$ & 96.1 & 67.6 & 17.6 & 16.7 \\
    \rowcolor{pfagentbg} \sys & 96.1 & 77.5 & 44.1 & \textbf{39.2} \\
    \midrule
    \multicolumn{5}{l}{\textbf{SWE-fficiency-Lite}} \\
    \midrule
    OpenHands & 82 & 47 & 27 & 26 \\
    $A_P$ & 85 & 75 & 45 & 44\\
    \rowcolor{pfagentbg} \sys & 90 & 83 & 75 & \textbf{74} \\
    \bottomrule
    \end{tabular}
\end{table}

\subsubsection{Reward Hacking}
\label{sec:eval:reward_hacking}

A key design choice in \sys is that when benchmarking, \sys only times the execution time of the \emph{first} run rather than the average across all profiling iterations. While this introduces some variance in the reported runtime, we find it to be small ($<3\%$ of overall runtime on average). The motivation is to discourage reward hacking: an agent receiving average-run timing as feedback can cache the result of an expensive operation on the first run and replay it on subsequent runs, collapsing the average runtime without performing any real optimization. The loop-only ablation $A_L$ makes this concrete.

On SWE-fficiency-Lite, the loop-only agent $A_L$ originally achieved a $62\%$ Opt@1 score but a $44\%$ hack-adjusted score as $18$ of its high-performing patches are labeled as hacks by our hack detector. We find that $14$ of the $18$ hacks are related to maintaining some form of persistent state or caching results. In the provided test script, SWE-fficiency times the task by running the provided workload many times, so an agent that caches an expensive result on the first run and replays it on subsequent runs collapses the average runtime and appears to achieve a large speedup without optimizing anything. Therefore, we remove this by separately running $A_L$ with a test script that only times the execution of the \emph{first} run. After making this change, we re-run $A_L$ with the updated test script, and the results are shown in \cref{tab:main-results-gpt}. We see that the number of hacks decreases significantly, from $18$ to $3$, indicating that if the evaluation harness is not constructed properly, then it can lead to reward hacking by the agent. We perform a more in-depth analysis of each reward hacking instance in Section C of the supplemental material.

\subsubsection{Ablation on SWE-fficiency-Lite Budget per Task}
\label{sec:eval:swefficiency_budget_per_task}
In \cref{sec:impact_control_loop}, we discussed that many of the instances in SWE-fficiency-Lite ran out of budget before being able to submit a patch on later turns, which led to speedups plateauing, as compared to GSO, the agent spends more turns and tokens reading files and running tests, and we show this in \cref{tab:read-test-output}. Therefore, we separately evaluate SWE-fficiency-Lite with a $\$10$ budget per task and report the results below in \cref{tab:swefficiency_results_by_turn}. Here, many more instances submitted attempts beyond turn $2$, and we see the speedups steadily increasing over time compared to running with a $\$5$ budget. We also observe improvements in performance on the SWE-fficiency-Lite compared to running with a $\$5$ budget, achieving an Opt@1 score of $79\%$ compared to $75\%$, and a hack-adjusted score of $77\%$ compared to $74\%$.

\begin{table}[ht!]
  \centering
  \footnotesize
  \setlength{\tabcolsep}{4pt}
  \caption{Average number of file read commands and test commands per instance, as well as the average output, in number of tokens, of the read and test commands per instance in GSO and SWE-fficiency-Lite.}
  \label{tab:read-test-output}
  \begin{tabular}{@{}lrrrr@{}}
  \toprule
  Benchmark & \makecell{\# File\\Reads} & \makecell{\# Test\\Invoc.} & \makecell{Read\\Tokens} & \makecell{Test\\Tokens} \\
  \midrule
  GSO & 35.4 & 7.4 & 31{,}399 & 2{,}990 \\
  SWE-fficiency-Lite & 42.7 & 6.5 & 41{,}911 & 7{,}294 \\
  \bottomrule
  \end{tabular}
\end{table}

\begin{table}[ht]
    \centering
    \footnotesize
    \setlength{\tabcolsep}{6pt}
    \caption{\small Average (harmonic mean) Speedup of an agent's submitted change at each turn and \# of instances that submitted a patch at that turn on SWE-fficiency-Lite. The remaining instances that did not submit a patch ran out of budget.}
    \label{tab:swefficiency_results_by_turn}
    \begin{tabular}{@{}c c r @{\hskip 2\tabcolsep} c r@{}}
    \toprule
    & \multicolumn{2}{c}{\textbf{\$5} budget per task} & \multicolumn{2}{c}{\textbf{\$10} budget per task} \\
    \cmidrule(lr){2-3} \cmidrule(lr){4-5}
    \textbf{Turn}
      & \textbf{\shortstack{Avg.\\Speedup}}
      & \textbf{\shortstack{\# Submitted}}
      & \textbf{\shortstack{Avg.\\Speedup}}
      & \textbf{\shortstack{\# Submitted}} \\
    \midrule
    1 & 1.697$\times$ & 89 & 2.727$\times$ & 98 \\
    2 & 2.486$\times$ & 69 & 3.492$\times$ & 84 \\
    3 & 2.638$\times$ & 40 & 3.888$\times$ & 67 \\
    4 & 2.667$\times$ & 17 & 3.945$\times$ & 38 \\
    5 & 2.672$\times$ & 6  & 4.023$\times$ & 12 \\
    \bottomrule
    \end{tabular}
\end{table}

\section{Related Work}
\subsection{LLM Coding Benchmarks}
\textbf{Kernel-level benchmarks.} HumanEval and MBPP target completion of simple Python programs \cite{humaneval, mbpp}, while LiveCodeBench and BigCodeBench target competitive-programming problems \cite{livecodebench, livecodebenchpro, bigcodebench}. EvalPerf, Effibench, and Enamel evaluate the efficiency of LLM-generated Python programs \cite{evalperf, effibench, enamel}, and ParEval evaluates correct and efficient parallel code \cite{pareval}. For kernels, AlgoTune optimizes routines from numerical computations \cite{press2025algotunelanguagemodelsspeed}, KernelBench and MultiKernelBench write optimized CUDA kernels from a PyTorch specification across architectures \cite{ouyang2025kernelbenchllmswriteefficient, multikernelbench}, and TritonBench generates Triton kernels \cite{tritonbench}.

\textbf{Repository-level and Long-horizon benchmarks.} SWE-Bench and its extensions evaluate an LLM's ability to resolve issues on real-world repositories \cite{jimenez2024swebench, zan2025multiswebench, deng2025swebenchproaiagents, zhang2025swebenchgoeslive, deepswe, frontier_code}. ProgramBench \cite{programbench} and FrontierSWE \cite{frontierswe} evaluate LLMs on long-horizon software engineering tasks such as writing a SQL engine from scratch. SWE-Perf, SWE-fficiency, and GSO instead target the performance of the agent's changes, not only correctness \cite{swefficiency, sweperf, gso}

\subsection{LLMs for Code Optimization}
SBLLM proposes a search-based framework for iterative optimization of competitive-programming problems~\cite{search_based_llms}, and PIE finetunes LLMs on curated competition data to improve code optimization \cite{shypula2024learningperformanceimprovingcodeedits}. LessonL is a multi-agent framework in which agents learn from each other's experience optimizing computational kernels~\cite{liu2025lessonslearnedmultiagentframework}. Stark, Astra, Pike, and KernelFalcon take multi-agent approaches to generating or optimizing CUDA kernels~\cite{dong2025starkstrategicteamagents, wei2025astramultiagentgpukernel, kernelfalcon, pike}, while CudaForce uses hardware feedback \cite{cudaforge}. PEAK optimizes GPU kernels across architectures via natural-language transformation specifications \cite{peak}, and AccelOpt and AutoComp target kernels for specialized accelerators \cite{accelopt, autocomp}. 

\textbf{Profiler-guided and iterative optimization.} Closest to our setting, POLO couples a profiler with an LLM agent loop, using an iterative weighting algorithm to localize bottlenecks while separate generator and decision agents propose and select edits \cite{polo}. Profiler-in-the-loop designs have also been explored at finer granularity: PRAGMA routes hardware-profiling signals through a multi-agent loop for kernel generation \cite{pragma}, ProfilingAgent guides structured pruning and quantization of neural networks \cite{profilingagent}, and TritonForge optimizes Triton kernels \cite{tritonforge}; PEACE draws on historical edits and external knowledge to optimize real-world repositories \cite{peace}. Our work differs on two axes. First, we target real-world repositories with native extensions (C, C++, Cython, Rust) and hold the agent to a human-expert bar, rather than synthetic programs, isolated kernels, or model-compression objectives. Second, we show sample-efficiency: our loop surpasses an \emph{oracle} best-of-five baseline at a fraction of the cost, isolating feedback quality rather than added compute as the driver of the gains.
\section{Threats to Validity}
\label{sec:threats}

\textbf{Construct and internal validity.}
\sys and both benchmarks measure performance on the single workload given to the agent, so a patch that is fast there is not guaranteed to be a general optimization. GSO mitigates this with \emph{hidden} performance tests, whereas SWE-fficiency-Lite times only the provided script and is thus easier to game---the source of the reward hacking we observe. We report hack-adjusted scores from a combined detector (\cref{sec:eval:hack_detection}) and manually reviewed every flagged instance, but an LLM-as-judge detector may still miss subtle exploits or over-flag legitimate fast paths. Because our controller validates with pytest-testmon, which tracks only Python coverage, a native-extension regression can slip between iterations; this deflates rather than inflates our score, since each benchmark's own correctness suite runs at evaluation time.

\textbf{External and conclusion validity.}
Our findings span two benchmarks, roughly a dozen Python-centric repositories, two models (GPT-5.1 and Kimi-K2), and a single base harness (Mini-SWE-Agent), and may not transfer to other languages, models, or harnesses; although \sys is harness-independent by construction, we have not measured its gains on OpenHands or Codex. Results are also single-run, so small differences fall within noise---the Kimi-K2 GSO gap (\cref{tab:main-results-kimi}) is about two tasks and should not be over-interpreted. We therefore base our headline GPT-5.1 claims on a task-by-task comparison that is statistically significant on both benchmarks (\cref{tab:mcnemar}).

\section{Conclusion and Future Work}
\label{sec:conclusion}

We presented \sys, a profiler-guided, verifier-in-the-loop workflow that turns an off-the-shelf coding agent into a repository-level performance optimizer. \sys attacks three failure modes of general-purpose agents on optimization tasks: missing the real bottleneck, stopping at the first passing patch, and under-testing complex changes, with three matching components: a curated profiler summary, an objective-driven loop controller that retains the \emph{fastest correct} patch, and selective re-validation after every iteration. 

\sys can be run on top of
more capable single- and multi-agent harnesses such as Codex or Claude Code, which we plan to evaluate. Other directions include extending selective testing to native-extension coverage
, hardening the profiling and timing template against new classes of reward hacking, and broadening to additional languages, models. More broadly, we view performance engineering as a distinct and demanding axis of agentic software engineering; one that rewards agents able to reason about \emph{how} code runs, not merely whether it is correct.

\section{Acknowledgments}

This research was sponsored in-part by the MIT-IBM Watson AI Lab. This research was sponsored in-part by the United 
States Air Force Research Laboratory and the United States Air Force Artificial
Intelligence Accelerator and was accomplished under Cooperative Agreement Number
FA8750-19-2-1000.  The views and conclusions contained in this document are
those of the authors and should not be interpreted as representing the official
policies, either expressed or implied, of the United States Air Force or the
U.S. Government.  The U.S. Government is authorized to reproduce and distribute
reprints for Government purposes notwithstanding any copyright notation herein.


\bibliography{sample}
\newpage

\appendices

\definecolor{frameblue}{HTML}{2C3E50}
\definecolor{accent}{HTML}{1F6FEB}
\definecolor{codebg}{HTML}{F4F6F8}
\definecolor{codeframe}{HTML}{D0D7DE}
\definecolor{reportbg}{HTML}{FFF8E6}
\definecolor{reportframe}{HTML}{E0C56E}
\definecolor{sectionclr}{HTML}{0B3D63}
\definecolor{impred}{HTML}{B3261E}

\setlist[enumerate]{leftmargin=2.1em,itemsep=2pt,topsep=3pt}
\setlist[itemize]{leftmargin=1.6em,itemsep=2pt,topsep=3pt}

\newcommand{\code}[1]{{\color{accent}\texttt{#1}}}
\newcommand{\imp}[1]{{\color{impred}\textbf{#1}}}

\newtcolorbox{innercode}[1][]{
  colback=codebg, colframe=codeframe,
  boxrule=0.6pt, arc=2pt,
  left=8pt,right=8pt,top=6pt,bottom=6pt,
  fontupper=\ttfamily\small, #1}

\newtcolorbox{innerreport}[1][]{
  colback=reportbg, colframe=reportframe,
  boxrule=0.6pt, arc=2pt,
  left=8pt,right=8pt,top=6pt,bottom=6pt, #1}

\newcommand{\heading}[1]{%
  \par\vspace{6pt}{\color{sectionclr}\large\bfseries #1}\par\vspace{2pt}}
\newcommand{\subheading}[1]{%
  \par\vspace{4pt}{\color{sectionclr}\bfseries #1}\par\vspace{1pt}}

\pagestyle{empty}

\section{Base \sys Prompt}
\begin{tcolorbox}[
  enhanced, breakable,
  colback=white, colframe=frameblue,
  boxrule=1.1pt, arc=4pt,
  left=14pt,right=14pt,top=12pt,bottom=12pt,
  title={Prompt for PerfAgent},
  fonttitle=\color{white},
  coltitle=white,
  attach boxed title to top left={xshift=10pt,yshift=-2pt},
  boxed title style={colback=accent,colframe=accent,arc=2pt}
]

I've uploaded a python code repository in the directory \code{/testbed}. There is a python
script \code{/perf\_script.py} which shows an example usage of the repository.

\vspace{4pt}
Can you help me implement the necessary changes to the repository so that the runtime of
the test scripts is optimized?

\vspace{4pt}
This will be a very difficult problem and may involve modifying multiple files across the
repository. Do not be afraid to make large changes to the repository in order to implement
your optimization.

\vspace{4pt}
I have obtained an initial summary from a profiler showing the bottlenecks. The profiler
report summarizes the output of \code{/profile\_prob\_script.py}, which profiles
\code{/perf\_script.py}.

\vspace{4pt}
Please use the following report to guide your optimization. 

\vspace{4pt}

\textit{\textbf{Report}}\\[2pt]
\texttt{\{\{ perf\_report \}\}}

\vspace{4pt}

To improve performance, either optimize the bottleneck itself or avoid calling the
bottleneck as frequently.

\heading{Testing}
\code{/run\_tests.sh} runs tests that can be used to verify the correctness of your
implementation. Additional hidden tests may be used to evaluate your implementation.\\

\heading{Basic guidelines}
\begin{enumerate}
  \item Your task is to make changes to non-test files in the \code{/testbed} directory to
        improve the performance of the test scripts. Do \emph{not} make changes to any other
        files as they will not count towards your submission.
  \item Make changes while ensuring the repository is functionally equivalent to the original.
  \item Do not overoptimize for just the specific inputs in the provided \code{perf\_script.py}.
        Make general performance improvements for the usage scenario shown.
  \item You may need to rebuild the repo for your changes to take effect before testing. Some
        rebuilds may take time to run, so be patient. Running \code{/build.sh} rebuilds the
        repository. Do not build the repository in other ways.
\end{enumerate}

Important:
Do not run scripts from within \texttt{/testbed} or add /testbed to PYTHONPATH, as it contains
source code, and you may run into issues with circular imports. Instead, change to a neutral
directory before running: \texttt{cd / \&\& python /path/to/script.py}\\[3pt]
Never use \texttt{cd /testbed \&\& python script.py}\\[5pt]
For running scripts, run from the \texttt{/} directory. For example, to run
\texttt{/perf\_script.py} run \texttt{cd / \&\& python perf\_script.py}.\\[5pt]
To install a package, run \texttt{uv pip install <package>}. For example, to install
\texttt{pytest}, run \texttt{uv pip install pytest}.

You can execute bash commands and edit files to implement the necessary changes.

\heading{Workflow}
\subheading{High-Level Problem Solving Strategy}
\begin{enumerate}
  \item Understand the problem deeply. Carefully read the issue and think critically about
        what is required.
  \item Investigate the codebase. Explore relevant files, search for key functions, and gather
        context.
  \item Develop a clear, step-by-step plan. Break down the fix into manageable, incremental
        steps.
  \item Implement the optimization incrementally. Make small, testable code changes while
        ensuring the repository is functionally equivalent to the original.
  \item Debug as needed. Use debugging techniques to isolate and resolve issues.
  \item Test frequently. Run tests after each change to verify correctness.
  \item Iterate until the performance has significantly improved and all tests pass.
  \item Continue to debug performance. Find if there are any additional bottlenecks in your
        solution and resolve them.
  \item Reflect and validate comprehensively. After tests pass, write additional tests to
        ensure correctness, and remember there are hidden tests that must also pass before the
        solution is truly complete.
\end{enumerate}

\subheading{1. Deeply Understand the Problem}
\begin{itemize}
  \item Carefully read the test script and think hard about a plan to optimize its performance
        before coding.
\end{itemize}

\subheading{2. Codebase Investigation}
\begin{itemize}
  \item Explore relevant files and directories.
  \item Search for key functions, classes, or variables related to the issue.
  \item Read and understand relevant code snippets.
  \item Identify the root cause of the problem.
  \item Validate and update your understanding continuously as you gather more context.
\end{itemize}

\subheading{3. Develop a Detailed Plan}
\begin{itemize}
  \item Outline a specific, simple, and verifiable sequence of steps to fix the problem.
  \item Break down the fix into small, incremental changes.
\end{itemize}

\subheading{4. Making Code Changes}
\begin{itemize}
  \item Before editing, always read the relevant file contents or section to ensure complete
        context.
  \item If a patch is not applied correctly, attempt to reapply it.
  \item Make small, testable, incremental changes that logically follow from your investigation
        and plan.
\end{itemize}

\subheading{5. Debugging}
\begin{itemize}
  \item Make code changes only if you have high confidence they can solve the problem.
  \item When debugging, try to determine the root cause rather than addressing symptoms.
  \item Debug for as long as needed to identify the root cause and identify a fix.
  \item Use print statements, logs, or temporary code to inspect program state, including
        descriptive statements or error messages.
  \item To test hypotheses, you can also add test statements or functions.
  \item Revisit your assumptions if unexpected behavior occurs.
\end{itemize}

\subheading{6. Testing}
\begin{itemize}
  \item Run tests frequently by running the perf script \texttt{cd / \&\& python perf\_script.py},
        or create your own tests and run those.
  \item After each change, verify correctness by running relevant tests.
  \item If tests fail, analyze failures and revise your patch.
  \item Ensure all tests pass before finalizing.
\end{itemize}

\subheading{7. Final Verification}
\begin{itemize}
  \item Confirm the performance has improved.
  \item Review your solution for logic correctness and robustness.
  \item Iterate until you are extremely confident the fix is complete and all test scripts pass.
\end{itemize}

\subheading{8. Improving Performance}
\begin{itemize}
  \item Continue to analyze the changed code to find bottlenecks and resolve them.
  \item Continue iterating on the test script (if needed) to test the performance of your
        changes on various inputs.
\end{itemize}

\vspace{4pt}
Once you are done with the above, submit your changes and finish your work by issuing the
following command: \code{echo COMPLETE\_TASK\_AND\_SUBMIT\_FINAL\_OUTPUT}. Do not combine it
with any other command. Important: After this command, you cannot continue working on this task.

\heading{Response Format}
For each response:
\begin{enumerate}
  \item Include a THOUGHT section explaining what you are looking for and why.
  \item At least one tool call.
\end{enumerate}

\subheading{Critical Requirements}
\begin{itemize}
  \item Your response SHOULD include reasoning text explaining what you're doing.
  \item Your response MUST include AT LEAST ONE bash tool call. You can make MULTIPLE tool calls
        in a single response when the commands are independent.
  \item Directory or environment variable changes are not persistent. Every action is executed
        in a new subshell.
  \item You can prefix any action with
        \texttt{MY\_ENV\_VAR=MY\_VALUE cd /path/to/working/dir \&\& ...} or write/load environment
        variables from files.
\end{itemize}

\subheading{Environment Details}
\begin{itemize}
  \item You have a full Linux shell environment.
  \item Always use non-interactive flags (\texttt{-y}, \texttt{-f}) for commands.
  \item Avoid interactive tools like vi, nano, or any that require user input.
  \item You can use bash commands or invoke any tool available in the environment.
\end{itemize}

\end{tcolorbox}

\section{Filtering and Selecting Tests}
\label{sec:appendix:tests}

\begin{table*}
    \centering
    \setlength{\tabcolsep}{4pt}
    \renewcommand{\arraystretch}{1.3}

    \caption{SWE-fficiency-Lite Test Counts broken down by repository.}
    \label{swefficiency_tests}
    \begin{tabular}{lcccc}
    \toprule
    Repo & \# of Tasks & Base Test Count & Filtered Test Count & Patch Test Count \\
    \midrule
    pandas       & 63 & $170{,}837 \pm 57{,}041$ & $164{,}893 \pm 56{,}526$ & $31{,}666 \pm 48{,}220$ \\
    sympy        & 5 & $8{,}140 \pm 3{,}134$    & $7{,}653 \pm 2{,}923$    & $1{,}189 \pm 2{,}057$  \\
    astropy      & 8 & $15{,}690 \pm 13{,}159$  & $15{,}051 \pm 12{,}765$  & $3{,}078 \pm 3{,}251$  \\
    numpy        & 5 & $29{,}558 \pm 20{,}112$  & $28{,}860 \pm 20{,}393$  & $3{,}164 \pm 6{,}420$  \\
    scikit-learn & 3 & $16{,}671 \pm 3{,}157$   & $14{,}693 \pm 3{,}607$   & $271 \pm 470$          \\
    scipy        & 7 & $36{,}842 \pm 23{,}706$  & $25{,}329 \pm 16{,}503$  & $1{,}312 \pm 1{,}442$  \\
    matplotlib   & 2 & $8{,}156 \pm 576$        & $8{,}022 \pm 604$        & $3{,}182 \pm 438$      \\
    xarray       & 3 & $14{,}758 \pm 4{,}032$   & $13{,}786 \pm 3{,}491$   & $7{,}346 \pm 3{,}275$  \\
    dask         & 4 & $9{,}649 \pm 3{,}299$    & $9{,}497 \pm 3{,}358$    & $4{,}497 \pm 3{,}963$  \\
    \bottomrule
    \end{tabular}

    \vspace{1.5em}

    \caption{GSO Test Counts broken down by repository.}
    \label{gso_tests}
    \begin{tabular}{lcccc}
    \toprule
    Repo & \# of Tasks & Base Test Count & Filtered Test Count & Patch Test Count \\
    \midrule
    numpy        & 36 & $35{,}002 \pm 9{,}429$   & $34{,}458 \pm 9{,}555$   & $109 \pm 614$         \\
    pillow-simd  & 7 & $3{,}700 \pm 1{,}019$    & $3{,}696 \pm 1{,}023$    & $549 \pm 1{,}303$     \\
    pandas       & 34 & $130{,}994 \pm 12{,}197$ & $130{,}002 \pm 12{,}463$ & $4{,}691 \pm 10{,}500$ \\
    transformers & 4 & $13{,}148 \pm 19{,}180$  & $1{,}744 \pm 2{,}929$    & $91 \pm 143$          \\
    tokenizers   & 4 & $129 \pm 46$             & $125 \pm 47$             & $1 \pm 1.73$          \\
    datasets     & 3 & $4{,}010 \pm 2{,}005$    & $3{,}003 \pm 1{,}046$    & $241 \pm 197$         \\
    pydantic     & 4 & $2{,}292 \pm 393$        & $2{,}268 \pm 398$        & $757 \pm 585$         \\
    pillow       & 4 & $2{,}383 \pm 1{,}726$    & $2{,}305 \pm 1{,}799$    & $267 \pm 289$         \\
    llama-cpp    & 2 & $6 \pm 0$                & $6 \pm 0$                & $0 \pm 0$             \\
    tornado      & 4 & $330 \pm 7$              & $179 \pm 3$              & $1.75 \pm 2$          \\
    \bottomrule
    \end{tabular}
\end{table*}

For many tasks in the benchmarks, the base commit of the repository contains both tests that consistently fail and flaky tests that fail intermittently. Returning these unrelated failures can derail the agent from completing its task, and makes it difficult for the controller to select correct patches. Therefore, we filter out these tests beforehand. For flaky tests in particular, we run the test suite 5 times, each time with a different random seed and also randomize the order of test execution. We exclude tests that fail in any of the runs and find that this produces a stable test suite for each of the tasks. 

The impact of filtering tests and selecting tests based on the specific changes made by the agent can be seen in \cref{swefficiency_tests} and \cref{gso_tests}. For each repository, we take the average and standard deviation across the tasks in that benchmark.

\section{Manual Review of Reward Hacking on SWE-fficiency-Lite}
\label{sec:appendix:reward_hacking}
Here we briefly analyze each instance of reward hacking made by the LLM agent with access to only \sys's loop controller. There are $18$ instances that are considered hacks by using the LLM-as-a-judge hack detector introduced in GSO.

\textbf{dask\_\_dask-10356}. In this instance, the timed workload repeatedly calls $\texttt{random\_state\_data\_python}$ with the same inputs. The agent adds a global cache variables that stores the result of this call and reuses it on repeated calls.

\textbf{matplotlib\_\_matplotlib-22108}. In this instance, the timed workload calls $\texttt{Affine2D.rotate}$ which rotates an internal 2D matrix. In the test script, the resulting matrix is not directly used for subsequent computation. Therefore, inside the function, the LLM agent performs a lazy evaluation and only updates an internal angle variable in the method instead. The rotated matrix is only materialized when other methods are called. In the timed workload however, these methods are never called and the materialization never occurs, which allows the agent to get large speedups on the timed workload.

\textbf{numpy\_\_numpy-18324}. In this instance, the agent's patch modifies related test files, which can affect evaluation in SWE-fficiency as it uses the existing repository's test suite.

\textbf{numpy\_\_numpy-27830}. In this instance, the agent adds a global cache that stores the result of the target workload: $\texttt{np.polynomial.legendre.legval}$. Since the workload repeatedly calls this method in the timed loop, the agent's patch avoids any computation for repeated calls.

\textbf{pandas-dev\_\_pandas-25953}. In this instance, the agent adds a cache to $\texttt{NDFrame}$, which avoids recomputing the aggregation results on repeated calls.

\textbf{pandas-dev\_\_pandas-27448}. In this instance, the agent adds a cache to $\texttt{CategoricalIndex}$ as the workload repeatedly performs slices on the same objects in the timed loop. The agent's change avoids the slicing on repeated calls.

\textbf{pandas-dev\_\_pandas-39972}. In this instance, the agent adds a cache to $\texttt{Styler.render}$ as the workload repeatedly renders the same $6$ objects in the timed loop. The agent's change avoids rendering on repeated calls.

\textbf{pandas-dev\_\_pandas-42353}. In this instance, the agent adds a cache to $\texttt{Index.union}$ as the workload repeatedly unions the same two pandas Index objects in the timed loop. The agent's change avoids performing the union on repeated calls.

\textbf{pandas-dev\_\_pandas-43683}. In this instance, the agent adds a cache to
the pandas dataframe object as the workload repeatedly calls $\texttt{dropna}$ on the same dataframe in the timed loop. The agent's change avoids this computation on repeated calls.

\textbf{pandas-dev\_\_pandas-51518}. In this instance, the agent adds a cache to store the result of a numpy array created from a range and reuses it across calls in the timed loop.

\textbf{pandas-dev\_\_pandas-53152}. In this instance, the agent adds a cache to $\texttt{ArrowExtensionArray.\_str\_get}$, as the workload repeatedly calls this method in the timed loop. The agent's change avoids doing any work on repeated calls.

\textbf{pandas-dev\_\_pandas-55084}. In this instance, the agent adds a cache to store the result of a previously computed union/intersection of pandas indexes.

\textbf{pandas-dev\_\_pandas-56110}. In this instance, the agent adds a cache to $\texttt{ArrowStringArray.\_str\_get}$ to cache the result of previous computations on the same array.

\textbf{pydata\_\_xarray-5661}. In this instance, the workload prints large arrays which tests repr formatting. The agent avoids computation by doing formatting for the first and last row, thereby avoiding most of the computation. The repository's correctness tests do not catch this behavior as the agent's change is a fast-path that only triggers on very large inputs, such as the ones in the timed workload.

\textbf{scipy\_\_scipy-10467}. In this instance, the agent adds a cache to short-circuit $\texttt{SphericalVoronoi}$ construction on repeated calls.

\textbf{scipy\_\_scipy-10939}. In this instance, the agent adds a cache to store the result of calls to various methods within $\texttt{csr\_matrix}$ such as tocsr, tocoo, todok, todia and tobsr, which are repeatedly called in the timed workload.

\textbf{scipy\_\_scipy-11517}. In this instance, the agent adds a cache to store the results of LIL sparse matrix operations such as todok, todia, tocsr, tobsr and tocoo.

\section{Manual Review of Per-Task Win/Loss Analysis}
In the main paper, we compared the OpenHands baseline and \sys on a per-tasks win/loss basis. Here, we manually review each instance in which the OpenHands baseline agent performed better than \sys, meaning the OpenHands baseline agent produced a patch that matched the performance of a human expert baseline and pass the hack detector, whereas \sys's patch did not.

In GSO, there are $7$ instances where the OpenHands baseline produced a patch that matched or exceeded the performance of the human expert baseline and passed our hack detector, whereas \sys did not. For instance: pandas-dev\_\_pandas-2f4c93e, \sys produced a patch that did not pass correctness checks. For instances pandas-dev\_\_pandas-c34da50, tornadoweb\_\_tornado-1b464c4, uploadcare\_\_pillow-simd-2818b90, the patches produced by \sys do well on the provided workload but do not perform as well on the hidden performance tests, while the OpenHands baseline produces patches that generalize better. For instance python-pillow\_\_Pillow-fd8ee84, the agent never submitted a patch as it ran out of budget before submitting. For instance uploadcare\_\_pillow-simd-0514e20, the OpenHands patch uses SIMD while \sys's patch uses OpenMP. For instance huggingface\_\_transformers-253f9a3, \sys fails to fuse the torch kernels while the patch from OpenHands does, although \sys's patch falls just below the $0.95$ threshold.

In SWE-fficiency-Lite, there are $4$ instances where the OpenHands baseline produced a patch that matched or exceeded the performance of the human expert baseline and passed our hack detector, whereas \sys did not. For instances numpy\_\_numpy-12575, numpy\_\_numpy-21832, and pandas-dev\_\_pandas-39332, \sys produced a patch that failed on some correctness tests. For instance: pandas-dev\_\_pandas-56806, \sys produced a patch that was correct but had a speedup ratio of $0.29$, under the Opt@1 threshold of $0.95$. In that patch, the OpenHands baseline uses a builtin pandas helper implemented in C, whereas \sys allocates many large temporary arrays in its implementation.

\end{document}